\documentclass[aps,pre,onecolumn,preprint,showpacs,showkeys,superscriptaddress,groupedaddress]{revtex4}

\usepackage{graphicx}
\usepackage{color}
\usepackage{subfigure}
\usepackage{dcolumn}   
\usepackage{amsmath}
\usepackage{float}
\usepackage{wasysym}
\usepackage{amssymb}
\usepackage{gensymb}
\usepackage{soul}
\usepackage{hyperref}
\newcommand{\be}{\begin{equation}}
\newcommand{\ee}{\end{equation}}
\newcommand{\bea}{\begin{eqnarray}}
\newcommand{\eea}{\end{eqnarray}}
\newcommand{\bse}{\begin{subequations}}
\newcommand{\ese}{\end{subequations}}

\newcommand{\tb}{\textcolor{black}}
\newcommand{\tr}{\textcolor{black}}
\newcommand{\tbb}{\textcolor{black}}
\newcommand{\trr}{\textcolor{black}}

\makeatletter
\newcommand*{\rom}[1]{\expandafter\@slowromancap\romannumeral #1@}
\makeatother

\begin{document}

\title{Mesoscale phase separation of chromatin in the nucleus}

\author{Gaurav Bajpai}
\affiliation{Department of Chemical and Biological Physics, Weizmann Institute of Science, Rehovot, Israel}
\author{Daria Amiad-Pavlov}

\author{Dana Lorber}
\author{Talila Volk}
\affiliation{Department of Molecular Genetics, Weizmann Institute of Science, Rehovot, Israel}
\author{Samuel Safran}
\affiliation{Department of Chemical and Biological Physics, Weizmann Institute of Science, Rehovot, Israel}

\email{sam.safran@weizmann.ac.il}
\email{gaurav.bajpai@weizmann.ac.il}

\date{\today}
\begin{abstract}
Intact-organism imaging of {\it Drosophila} larvae reveals and quantifies chromatin-aqueous phase separation.  The chromatin can be organized near the lamina layer of the nuclear envelope, conventionally fill the nucleus,  be organized centrally, or as a wetting droplet.  These transitions are controlled by changes in nuclear volume and the interaction of chromatin with the lamina (part of the nuclear envelope) at the nuclear periphery. \trr{Using a simple polymeric model that includes the key features of chromatin self-attraction and its binding to the lamina, we demonstrate theoretically that it is the competition of these two effects that determines the mode of chromatin distribution}. The qualitative trends as well as the compositional profiles obtained in our simulations compare well with the observed intact-organism imaging and quantification. Since the simulations contain only a small number of physical variables we can identify the generic mechanisms underlying the changes in the observed phase separations.

\end{abstract}

\maketitle

\section*{Introduction}
Chromatin is a complex, linear macromolecule comprising DNA and histone proteins which in eukaryotic cells, is localized in the nucleus where it is solubilized in water, salts and other small molecules~\cite{cooper2000molecular,phillips2012physical}. In many studies, chromatin organization in interphase is homogeneous on the nuclear scale. In this ``conventional'' picture, the chromatin and the aqueous solvent uniformly fill the nucleus as a single phase ~\cite{rosa2013insights}.  However, even the conventional picture accounts for phase separation similar to that of soluble AB block copolymers~\cite{rubinstein2003polymer}, with regions of transcriptionally active eurchromatin (A block) separated from regions of relatively inactive heterochromatin (B block); however both are assumed to be homogeneously solubilized in the aqueous phase~\cite{erdel2018formation,narlikar2020phase,Strom2017,Larson2017}. For example, Hi-C experiments reveal such AB  chromatin comparmentalization ~\cite{Lieberman-aiden2009} \tr{ but do not provide information about their location within the nucleus}. 

In most previous studies, nuclear-scale phase separation of the chromatin and the aqueous phase has not been considered.  However, recent super-resolution microscopy reveals a ``marshland'' of chromatin and aqueous phase, with a non-uniform distribution of chromatin at submicron level~\cite{cremer20154d}. Another study observed a larger-scale phase separation of chromatin and the aqueous phase in early development, with the chromatin localized to the nuclear periphery~\cite{Popken2014}. Both these observations did not distinguish the A and B blocks (eu and hetero chromatin) and instead identified large regions of DNA-rich chromatin separated from DNA-poor regions, presumably of the aqueous phase. More recently, \tr{intact-organism} imaging of {\it Drosophila} larvae (where the chromatin was labeled by H2B-RFP, and the nuclear envelope was labeled by Nesprin/Klar-GFP) reveals and quantifies chromatin-aqueous phase separation and its control by changes in the nuclear volume and the interaction of chromatin with the lamina (part of the nuclear envelope) at the nuclear periphery~\cite{Amiad-pavlov2020}. Here, we demonstrate theoretically that it is the competition of these two effects, together with the self-attraction of the chromatin to itself that determine whether the chromatin is conventionally or peripherally distributed within the nucleus; we also show that more complex organizational modes  are also possible. The model we present below that focuses on the concentration profile of chromatin {\it(whether A or B)} within the nucleus, is appropriate when the chromatin self-attraction compared with the chromatin-aqueous phase interaction as well as the chromatin-lamina interaction are larger than the difference between the AA or BB interactions. The qualitative trends as well as the compositional profiles obtained in our simulations compare well with the observed \tr{intact-organism} imaging and quantification. Since the simulations contain only a small number of physical variables, this allows us to identify the generic mechanisms underlying the changes in the phase separation that are observed.

The physical insight obtained here focuses on the competition of two primary interactions of chromatin within the nucleus: (i) chromatin-lamina, which is attractive and tends to organize the chromatin peripherally (ii) chromatin-chromatin, that if attractive, tends to condense the chromatin and separate it from the aqueous phase. The nuclear envelope contains lamina which is a dense, fibrillar network of proteins that provide an anchoring points for various proteins that bind to chromatin on one end and the lamina on the other~\cite{ulianov2019nuclear,Kind2013,Yanez-Cuna2017a}. These binding proteins associate with lamins that in general comprise two A-type (lamin A and lamin C) and two B-type (lamin B1 and lamin B2) proteins~\cite{moir2000nuclear}. \trr{However, {\it Drosophila} has only one A-type lamin (lamin C) and one B-type lamin (lamin Dm0) gene~\cite{riemer1995expression,schulze2009comparative}.} A-type and B-type lamins are found at the nuclear periphery while A-type lamins are also found in the nuclear interior~\cite{Briand2020,naetar2017lamins}. Those chromatin domains, which bind to the lamins are called lamina associated domains (LADs) which are similarly divided in two groups: A-LAD and B-LAD corresponding to interactions with A-type lamin and B-type lamin respectively~\cite{VanSteensel2017,Briand2020}. The anchor protein lamin B receptor (LBR) binds lamin B and acts as a tether between chromatin and the nuclear lamina~\cite{Solovei2013}. Experiments performed in the absence of LBR observed a loss of peripheral LADs and an inverted architecture with LADs localized within the nuclear interior~\cite{Solovei2013,Briand2020}. \trr{The role of lamin Dmo in {\it Drosophila} is the same as LBR in mammalian cells~\cite{wagner2004lamin,ulianov2019nuclear}}.

Previous models of chromatin-lamina interactions resolved euchromatin and heterochromatin domains in the context of conventional organization of chromatin~\cite{Chiang2019} where phase separation with the aqueous phase is not considered. Another model of chromatin and lamina interactions, treats chromatin as a self-avoiding polymer in good solvent and focuses on the role of lamina in the formation of chromosome territories ~\cite{Maji2020}.  \tr{ However, {if} chromatin were a self-avoiding polymer (with no self-attraction), its radius of gyration in aqueous solution would typically be larger than {the} diameter of the nucleus.  Thus, under confinement, chromatin would fill the entire volume of nucleus, which is not consistent with the 
observation~\cite{Amiad-pavlov2020} of peripheral or central chromatin organization. }In contrast, the study presented here illustrates in an intuitive manner, how chromatin-lamin interactions control peripheral and other organizational modes that show chromatin-aqueous phase separation. The experimental observations of peripheral organization demonstrate the condensation of chromatin in the outer part of the nucleus, indicating the presence of chromatin self-attraction~\cite{Amiad-pavlov2020}. This motivates the model which focuses on attractive interactions within the chromatin so that it acts in the nucleus as a polymer in a relatively bad solvent.  For large enough nuclear volumes and strong enough chromatin-lamina attractions, instead of filling the nucleus, the chromatin is condensed in only part of the volume near the nuclear periphery.

The model we study is generic in nature, short-ranged and does not depend on the detailed molecular origin of the chromatin self-attraction.  This is an appropriate approach for understanding and predicting the nuclear-scale concentration profile of the chromatin and aqueous regions in-vivo, where the identity and function of all the molecular actors are so far unknown.  Nevertheless, there is ample reason to believe that such attractions are present and important in determining chromatin organization. For example, the positively charged histone tails attract negative DNA linkers and can thus promote condensation of chromatin~\cite{Rosen2019,Bajpai2019,hancock2007packing}. Protein condensates that interact with chromatin can also result in self-attraction.  A prominent example is the phase separation of HP1 in the nucleus, which binds to the heterochromatin domains of chromatin ~\cite{Strom2017,Larson2017}.  If one protein can bridge more than one chromatin regions that are close in physical space, but possibly far along the stretched chromatin chain, this can also result in an effective chromatin-chromatin attraction. Other nuclear proteins that interact with chromatin can play a similar role.  For example, nucleoplasmic lamin A can act to ``crosslink'' two chromatin regions in vivo~\cite{Bronshtein2016,Bronshtein2015,Marko2017}. Other studies have indeed considered chromatin to act as a polymer gel cross-linked by non-histone protein complexes~\cite{Marko2019,Discher2013}.

\section*{Model and Method}
As explained above, we are interested in the {\it generic} physical effects that determine the chromatin concentration profile at the nuclear scale.  However, to predict this from computer simulations of chromatin as a self-attracting, confined polymer that is also attracted to the lamina at the nuclear periphery, we must treat a specific polymeric model. We therefore fixed the polymer monomer size, molecular weight, and persistence length, and instead varied the physical properties of the nucleus including its volume and chromatin-lamin interactions.  This was motivated by the \tr{intact-organism} 
experiments~\cite{Amiad-pavlov2020}.  Comparing the live to fixed cells showed a reduction in  the nuclear volume by a factor of about 3 and a transition from peripheral (live) to conventional (fixed) chromatin organization.  Lamin C overexpression resulted in a transition from peripheral (WT) to central (and wetting droplet) organization in the mutant. In our simulations, we also varied the magnitude of the chromatin self-attraction to demonstrate the difference in organization of chromatin in a good solvent vs. a poor one.

{\bf Chromatin chain:} For the understanding of the generic properties of interest here, we treated a single chromosome.  Aqueous phase separation in case of multiple chromosomes can be mapped from our results by using the appropriate chromatin volume fraction.  We were motivated by the X chromosome of {\it Drosophila} since its distribution of LAD domains is known.  We therefore modeled chromatin as a polymer using a bead-spring model~\cite{Naumova2013}  with $N=37,333$ beads. Each bead represents 3 nucleosomes (around 600bp of DNA) whose diameter is denoted by  $\sigma=10$~nm. Thus, our model chromosome contains about 22.4Mbp, comparable to the X chromosome.  The bead-spring interactions ($E^{\textrm{chrom}}_s$) and persistence length (bending energy $E^{\textrm{chrom}}_b$) are determined in the standard manner as detailed in the { Supplementary Information (SI) text}, where it is shown that the appropriate persistence length is 2 beads in our representation.

In addition to the spring-bead model that accounts for the connectivity of nearest neighbor beads along the polymer chain, we include short-range interactions (both attractive and repulsive) between {\it any} two beads that are close enough in 3d space; they do not necessarily have to be ``close'' along the chain. These, non-bonding interactions between any two beads were taken using Lennard-Jones (LJ) potential,
\be
 E_{LJ}= \left\{
  \begin{array}{l l}
    4\epsilon\sum \limits_{\substack {{i,j}\\{i<j}}}\left[\left(\frac{\sigma}{r_{ij}}\right)^{12}-\left(\frac{\sigma}{r_{ij}}\right)^{6}\right] & \quad \text{when ${r}_{ij}<r_c$ }\\
    0 & \quad \text{when ${r}_{ij}\geq r_c$}
  \end{array} \right.
\ee

where $r_c$ refers to a cutoff distance beyond which LJ interaction is set to zero.  Here ${r}_{ij}$ is the distance in 3d space between $i^{th}$ and $j^{th}$ beads and $\epsilon$ is strength of potential. If there are no attractive forces between the beads (but only steric, hard-core repulsions), we truncate the LJ potential at the distance at which the repulsive force goes to zero which gives $r_c=2^{1/6}\sigma$.  For the case of attractive forces, of particular interest in our work, the cutoff distance is taken as $r_c=2.5\sigma$. For $\epsilon=1 k_BT$, the chromosome is a self-avoiding chain when $r_c=2^{1/6}\sigma$ and a self-attractive one when $r_c=2.5\sigma$ (see {SI text} and Fig.~S1 for scaling exponent results). 

{\bf Nuclear confinement and volume:} We model the nucleus to which the 
chromatin is  localized, as spherical shell of confinement radius $R_c$ and 
define a confinement potential $E^{wall}_{LJ}$ that accounts for the hard-core 
repulsion between the chromatin beads and the spherical wall. Each bead of the 
chromatin chain interacts with its nearest point on the wall through the 
potential $E_{LJ}$ with LJ strength $1k_BT$ and the nuclear-chromatin cut off 
distance $r_n=2^{1/6}\sigma$.  This is the distance at which the repulsive 
force between the chromatin beads and the nuclear shell falls to zero.  
Motivated by the experiments where the nuclear volumes for live nuclei are 
significantly larger than those of fixed nuclei, we allow for variations of the 
confinement volume (with fixed chromatin volume fixed $N$ and bead size). 
\tb{We define the parameter $\phi$, as global volume fraction of chromatin 
within nucleus, with $(1-\phi)$, the volume fraction of the aqueous phase}:
\be
\phi=\frac{\text{Volume of chromatin chain}}{\text{Volume of confinement}}=\frac{N \times \frac{4}{3}\pi (\sigma/2)^3}{\frac{4}{3}\pi R_c^3}
\ee

 We varied $R_c$ and hence $\phi$ from 0.1 to 0.5, where small value of $\phi$ models the case of chromatin in hydrated nuclei (aqueous solution) and large value of $\phi$ is for chromatin in dehydrated nuclei. 

{\bf Chromatin LAD and non-LAD domains:} To account for the biophysics of the bonding of the LAD domains of chromatin to the lamina, our chain comprises two types of beads in which only the LAD beads can be bonded to the lamina. The remaining non-LAD beads do not form such bonds. In order to simulate a specific system, we modeled the the 22.4 Mbp regions of chromosome X (ChrX) of {\it Drosophila} where the distribution of the LAD regions is available online~\cite{ho2014comparative}. That data indicates that 48\% of the sequences are LAD with an average domain size  of 90 kbp. We analyzed the data which shows an alternating patterns of LAD and non-LAD along the chromosome length (see Fig.~S2 in the SI) and used these patterns in our simulations. To demonstrate that our coarse-grain model does not depend on the size of the chromosome and the locations of the LADs along the chromosome, we also used a Monte-Carlo method to randomly distribute the LAD regions along the chain (see details of method and results in SI text and Fig.~S3 \& Fig.~S4).

{\bf Chromatin-chromatin interactions}
At the microscopic level, there are many possible interactions (e.g., electrostatic attraction between DNA and histone-tails, phase separation of  the DNA binding non-histone protein HP1, and DNA crosslinking by nucleoplasmic lamin-A) which can cause chromatin self-attraction and lead to its separation from the aqueous phase. Since there may be additional bridging proteins and because even the ones we listed are not known in detail, we use a generic model for short-range chromatin-chromatin attractions.  As outlined above, we model chromatin self-attraction by an  attractive LJ potential (with cutoff $r_c=2.5\sigma$ as explained above). In the section below that considers variations of the interaction strength $\epsilon$, we first consider the case where all beads (both LAD and non-LAD) interact with the same LJ potential. 
%

{\bf Chromatin-lamina interactions}
To model the lamina that interacts with the LAD domains of the chromatin, we introduce additional beads which are localized to the
confinement surface. These lamina beads are static in our simulation either as a simple mean-field approximation of the lamina or possibly
because the dynamics of the  nuclear lamina (NL) are slower than chromatin dynamics. The very short-range LAD-lamina bonding is mediated by specific proteins which we account for below. However, for the LAD regions of the chromatin to find the lamina within our simulations, we include an additional, somewhat longer-range attraction, modeled by a relatively weak, LJ potential with $\epsilon_{lm}=1k_BT$ with cutoff of $r_c=2.5\sigma$ as explained above, where $\sigma$ is the minimum hard-core distance between LAD and lamina beads.

The interactions of the non-LAD beads with the lamina are purely repulsive (due to their excluded volume) and to model that we use a LJ interaction with $\epsilon_{ln}=1k_BT$ and cutoff distance $r_c=2^{1/6}\sigma$, at which point the repulsive force falls to zero. The total, non-bonding (physical) interaction energy between lamina and chromatin ($E^{\textrm{lamina}}_{LJ}$) is calculated from the sum of these individual LAD and non-LAD LJ interactions with lamina.

However, the most important feature of the chromatin-lamina interaction is the bonding of the LAD and the lamina via specific proteins.  This depends on the availability of these proteins and \tr{the nature of their binding to the} lamina. In the experiments~\cite{Amiad-pavlov2020}  that induced lamin C overexpression, there were significant changes in the chromatin organization with much less bonding of the chromatin to the lamina.  Lamin overexpression has been suggested to repress the bonding activity of LBR proteins that bind the LAD domains to lamin B~\cite{buxboim2017coordinated,cho2017mechanosensing}. We therefore allow for the possibility that not all LAD domains can bind to the lamina by introducing the parameter $0 \leq \psi \leq 1$ which represents the fraction of the LAD domains that can bind to the lamina {(for example, due to a reduction in the number of binding proteins relative to the number of LAD)}.  

When  $\psi=1$ all LAD can bond to the lamin beads while $\psi=0$ represents the situation in which no LAD-lamina bonds are possible (e.g., due to a lack of bonding proteins or bonding sites within the lamina). We allow for bond formation when a LAD bead is within a distance of $r_b=2.5\sigma$ from a lamin bead where the LJ interaction is attractive.  A bond is then formed with an energy $U_b=-k_{bond}(r_b-\sigma)^2$, where we take $k_{bond}=10k_BT/\sigma^2$. The energetic cost of changing the bond length $r$ from an optimal value of $r=\sigma$, is accounted for by  introducing a  stiff spring connecting the LAD and lamin beads with the spring constant, $k_{bond}=10k_BT/\sigma^2$ and equilibrium spring-length of $\sigma$.  Thus, when the LAD-lamin distance is equal to $\sigma$, the spring energy and the LJ interaction are both zero and the optimal bond energy, $U_b$, is attained.  When the distance is either larger or smaller than $\sigma$, the energy due to the stretching of the spring is positive; at distances smaller than $\sigma$, the LJ repulsion dominates due to the excluded volume.  In our model, a single LAD bead can form a maximum of one bond with a given lamin bead. An existing bond can also break with a probability ($P_{break}=\exp{(-k_{bond}/k_BT)})$, when the difference of the LAD-lamin bead distance and is larger than $r_b$. (In the SI text, we describe variations in the the cutoff distance for bond breaking.) We used a kinetic Monte Carlo method to create/break bonds between the lamin and LAD beads. The total energy of LAD-lamin bond creation or breaking is defined by the spring potential $E^{\textrm{lamina}}_s$.

Adding up all these contributions yields the total potential energy of chromatin in our model nucleus:
\be
E_{tot}= E^{\textrm{chrom}}_s+E^{\textrm{chrom}}_b+E^{\textrm{chrom}}_{LJ}+E^{\textrm{wall}}_{LJ}+E^{\textrm{lamina}}_{LJ}+E^{\textrm{lamina}}_s
\ee

The total potential energy depends on the positions of the beads that determine their mutual interactions and those of the LAD beads with the lamina; the force on a given bead arises from the gradient of this potential energy.  The system is simulated calculating the Brownian dynamics of the chromatin beads in which the aqueous solvent is treated implicitly by the average frictional force it exerts on the moving beads and by the deviations from this average, represented by a stochastic force.  For the use in computer simulations, the discrete time~\cite{Doyle2005,LAMMPS1995}  equation of motion of the beads in this model is given by  the Langevin equation:
\be
 {\bf r}_i(t+\Delta t)={\bf r}_i(t)-\frac{\Delta t}{\gamma m}\nabla_{{\bf r}_i} E_{tot}(t)+\sqrt{\frac{6 k_BT\Delta t}{\gamma m}}\mbox{\boldmath$\xi$}_i(t)
\ee
where $m$ is bead mass, $\Delta t$ is the time-step, $\gamma$ is damping contact (related to the friction of the beads with the aqueous solvent) and {\boldmath$\xi$} is the stochastic force (thermal noise applied by the implicit aqueous solvent to each bead) and has the standard statistical properties used in Brownian dynamics simulations of equilibrium systems~\cite{Doyle2005}. The LAMMPS package is used for our calculations~\cite{LAMMPS1995}.

{\bf Initial conditions and equilibration:} Our initial condition has the chromatin chain comprising both LAD and non-LAD domains (that are many sequences long) that are mixed along the chain as indicated by the sequence data~\cite{ho2014comparative}. The center of mass of the chain is initially at the center of the nucleus and the dynamics are determined by the numerical solution of the stochastic Langevin equation that in general, includes the forces that arise from the interactions of the chromatin beads with themselves and the LAD domains with the lamina, plus the stochastic forces due to the implicit solvent. Initially, we include only the excluded volume chromatin-chromatin interactions, LAD-lamin attractions with LJ interactions and allow a fraction $\psi$ of LAD domains within a distance $r_b$ of the lamina  to create/break bonds with lamin beads,  via the spring model, allowing for both bond creation and breaking dynamics. After equilibration, we changed the chromatin-chromatin interactions to allow for self-attraction and let the system re-equilibrate, again using the LAMMPS package~\cite{LAMMPS1995}. The parameter values used in the simulations are summarized in Table~1 of the SI.

%

\section*{Results}

We now summarize the results of Brownian dynamics simulations and discuss how conventional, peripheral and central organization of chromatin can be achieved by changing hydration (chromatin volume fraction $\phi$), chromatin-lamina interactions  (fraction of bonded LAD $\psi$), and intra-chromatin interactions (attraction strength $\epsilon$).  The simulations below demonstrate and quantify how the competition of chromatin self-attraction, hydration, and LAD-lamina attraction compete to determine the qualitatively different chromatin concentration profiles shown below. This is summarized at the end of this section in a state diagram showing the transitions between the various nuclear-scale chromatin organizational modes.  The comparison with the experimental trends is addressed in the Discussion section.

\begin{figure} [H]
\begin{center}
\includegraphics[scale=0.3]{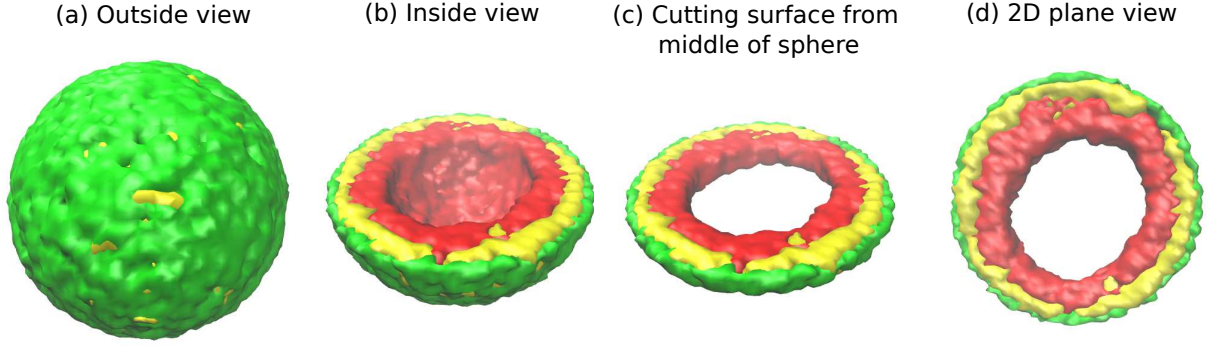}
\caption{{\bf Snapshots of the simulated system from different views}
        This representation is generated by drawing the surfaces around the beads that represent the lamin and chromatin. The same set of pictures shown here are depicted in a different graphical representation, in terms of the beads themselves, in Fig.S5 in the SI. (a) {\it Outside view:} Our model, spherical nucleus is enclosed by lamin (NL) beads (green). Within the sphere, the chromatin chain of N=37,333 beads contains two types of beads: LAD (yellow) and non-LAD (red). (b) {\it Inside view:} The sphere is cut at the equatorial plane to reveal one  hemisphere, so that chromatin (red and yellow) is visible. The central region of the nucleus is devoid of chromatin for the particular conditions ($\phi=0.3, \psi=1, \epsilon=1$) of this simulation. (c) {\it Cutting a slice near the equatorial plane:} A slice is cut from the hemisphere, resulting in a  3D surface which has width of $(1/10)th$ of the sphere diameter. (d) {\it 2D planar view:} We show the equitorial (xy) plane view of the 3D slice and observe that the central regions contains no chromatin.
\label{fig:01}
 } 
\end{center}
\end{figure}

{\bf Peripheral organization of chromatin:} Before presenting the transitions in chromatin organization as a function of the parameters discussed above, we first discuss one important example of peripheral organization in pictorial detail. Motivated by the experiments, we tune the simulation parameters $\phi, \psi$ and $\epsilon$ to the range where the equilibrium organization of the chromatin is  peripheral, for example:  $\phi=0.3, \psi=1, \epsilon=1$. In Fig.~\ref{fig:01}, we present snapshots of simulations in which the (a) {\it outside view} depicts the outer surface of nuclear lamina (green). To see inside, we cut the spherical system along its equatorial plane. (b) In the {\it inside view}, we can see the phase separation of chromatin and the aqueous phase, where the LAD regions of the chromatin (yellow) are in contact with the nuclear lamina (green) and the non-LAD regions (red) are largely separated from the LAD regions (yellow). We next show a  slice of this hemisphere including the equatorial plane in  Figs.~\ref{fig:01} (c) and (d), demonstrating  that there is no chromatin in the central part of the ring and that it contains the (implicit) aqueous solvent. Thus, Fig.~\ref{fig:01}, demonstrates that for these particular values of $(\phi,\psi,\epsilon)$ the chromatin is  localized near the nuclear periphery. We next systematically vary these important physical parameters and show how the simulation results change to reveal different organizational modes of the chromatin.

\begin{figure} [H] 
\begin{center}
\includegraphics[scale=0.22]{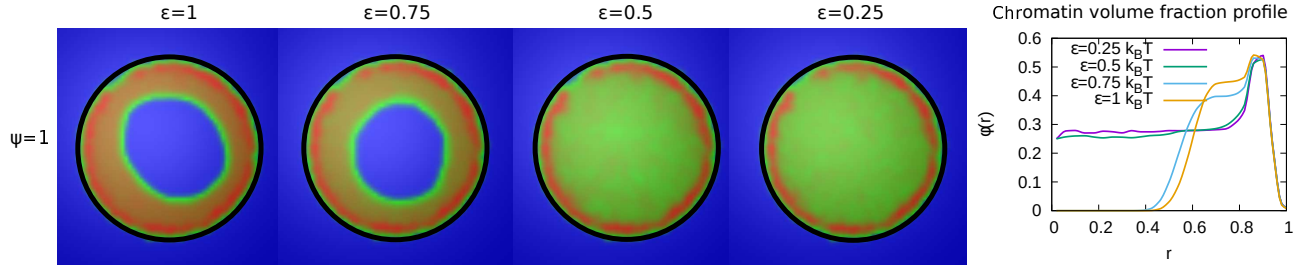}
\caption{{\it Left panel:} Chromatin concentrations are shown for different 
intra-chromatin attraction strengths ($\epsilon$) with a volume fraction of 
chromatin $\phi=0.3$ and maximal LAD-lamina interactions ($\psi=1$). For 
smaller values of the attractions, the chromatin no longer shows peripheral 
localization, and fills the entire nucleus.  This demonstrates the role of the 
chromatin self-attractions in stabilizing peripheral chromatin organization 
with a relatively high local volume fraction (red) of chromatin compared to the 
case of relatively small self-attractions where the chromatin fills the entire 
nucleus, with a smaller local volume fraction (green).\tb{ {\it Right panel:} 
Radial local volume fraction profiles of chromatin  for different ${\epsilon}$ 
plotted as a function of the radial distance $r$ where $r=0$ is the nuclear 
center and $r=1$ is the position of the nuclear envelope.}
\label{fig:02}
 } 
\end{center}
\end{figure}

{\bf Variation of intra-chromatin attractive interactions:}  
In our simulations, 48\% of the chromatin are LAD domains (corresponding to  
$\psi=1$) which remain strongly bound to the lamina. For a chromatin volume 
fraction $\phi=0.3$ and maximal chromatin-lamina attraction (i.e., all LAD 
domains can bond to the laminar, $\psi=1$), we varied the value of the 
chromatin self-attraction, $\epsilon$, which is the same for all pairs of beads 
(LAD pairs, non-LAD pairs, and LAD/non-LAD pairs). The LAD-lamin interaction 
strength and the chromatin volume fraction were held fixed.  In the left panel 
of Fig.~\ref{fig:02}, we depict the chromatin concentrations  in the 
equatorial, xy-plane calculated by taking average of many  frames (snapshots), 
while the system remained in equilibrium. The chromatin concentrations are 
shown with different colors where blue is chromatin-free and thus represents 
the aqueous phase, green represents low chromatin concentrations and red, high 
concentrations. When the intra-chromatin attractive interaction strength is 
relatively large, $\epsilon=1$, the chromatin organization is peripheral. When 
the self-attraction is sufficiently small ( $ \sim \epsilon <1/2$), the 
chromatin fills the entire volume, showing conventional organization. \tb{In 
the right hand panel of Fig.~\ref{fig:02}, we plot the local volume fraction of 
chromatin $\phi(r)$, within a spherical shell as a function of the normalized 
radial distance $r$, where $r=0$ represents the nuclear center and $r=1$ 
represents the position of the nuclear envelope.  We show the results for 
different values of the chromatin self-attraction.  Here, $\phi(r)$ is the 
local volume fraction of chromatin; the and average  of $\phi(r)$ over the 
spherical volume is the global volume fraction $\phi$. In the plot, we observe 
that for higher intra-chromatin attraction (${\epsilon}\geq0.75$), the local 
volume fraction shows a peak near the nuclear periphery ($r=1$) and decreases 
to zero near the center, consistent with peripheral organization. For smaller 
intra-chromatin attraction (${\epsilon}\leq0.5$), the local volume fraction} 
has a peak near periphery ($r=1$) but never decreases to zero since, in 
conventional organization, the chromatin fills the entire volume of the 
nucleus. This demonstrates the role of the chromatin self-attractions in 
stabilizing peripheral chromatin organization with a relatively high local 
volume fraction (red) of chromatin compared to the case of relatively small 
self-attractions where the chromatin fills the entire nucleus, with a smaller 
local volume fraction (green).

\begin{figure} [H] 
\begin{center}
\includegraphics[scale=0.25]{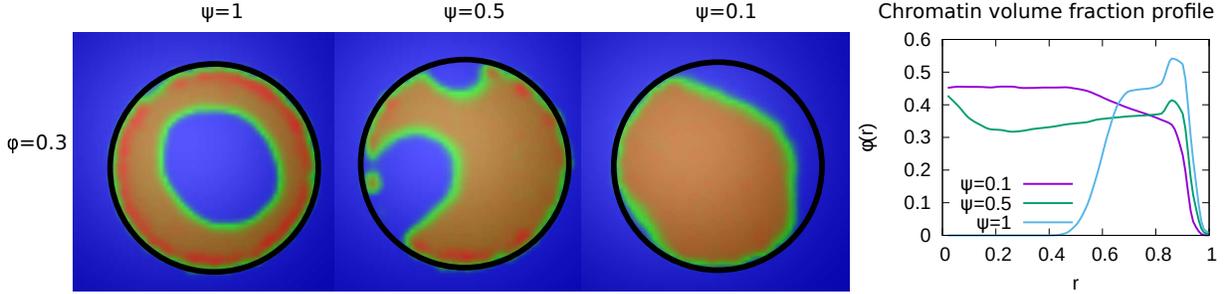}
\caption{{\it Left panel:} Variation in the fraction $\psi$ of LAD beads that can bond to the lamin, associated domains relative to its maximal value of $\psi=1$, where 48\% of the chromatin consists of LAD domains that can bind to the lamin. The values of the chromatin volume fraction $\phi=0.3$ and chromatin self-interaction strength, $\epsilon=1$ are fixed.  Those LAD domains (a fraction, $1-\psi$) not bound to the lamin are not necessarily found near the periphery of the nucleus and are mixed with the non-LAD chromatin.  For small values of $\psi$ the chromatin is no longer peripherally localized but fills the nucleus more uniformly (but see the discussion of the ``wetting droplet'' in the SI text and Fig.~S7), since there are relatively few LAD-lamin bonds to localize the chromatin at the nuclear periphery.  
 {\it Right panel:} Radial density plots of chromatin (each normalized to their maximal values) as a function of the radial distance $r$, where $r=0$ is the nuclear center and $r=1$ is the location of the nuclear envelope.
\label{fig:03}
 } 
\end{center}
\end{figure}

{\bf Variation of the fraction of the LAD domains bound to the lamina:} In our 
simulations, we considered the case in which not all LAD domains (in our 
simulations, 48\% of the chromatin) can bind to the lamin beads. This was 
suggested by the experiments in systems where lamin C was overexpressed and may 
be due to a reduction in the number of lamin binding proteins or the obscuring 
or burial of lamin binding sites in the overexpressed 
situation~\cite{buxboim2017coordinated}. The parameter $\psi$ represents the 
fraction of LAD beads that can potentially bind to the lamin.  We now present 
the results of simulations that vary the value of $\psi$ for fixed values of  
$\phi=0.3$ and ${\epsilon}=1$.  The left hand panel of Fig.~\ref{fig:03}, shows 
that a  decrease in the value of $\psi$ results in a transition of the 
chromatin organization from peripheral to central.  While in the equatorial 
plane, this is indeed  mostly central organization, the 3D structure is more 
complex with a ``droplet" of chromatin which ``wets'' and thus contacts the 
bottom of the hemisphere due to the LAD-lamin attractions.  This ``wetting 
droplet" configuration is discussed in more detail and shown in the SI text and Fig.~S7, where 
its kinetic features are also explored.   A similar progression is observed for 
a small chromatin volume fraction, $\phi=0.1$, and shown in Fig.~S6 in the SI. In the 
right hand panel of Fig.~\ref{fig:03}, the \tb{local chromatin volume 
fraction}  shows a peak near the nuclar periphery ($r = 1$) for $\psi = 1$ 
while for $\psi=0.1$ the peak of the \tb{local volume fraction} is shifted 
towards the center ($r=0$).

\begin{figure} [H] 
\begin{center}
\includegraphics[scale=0.22]{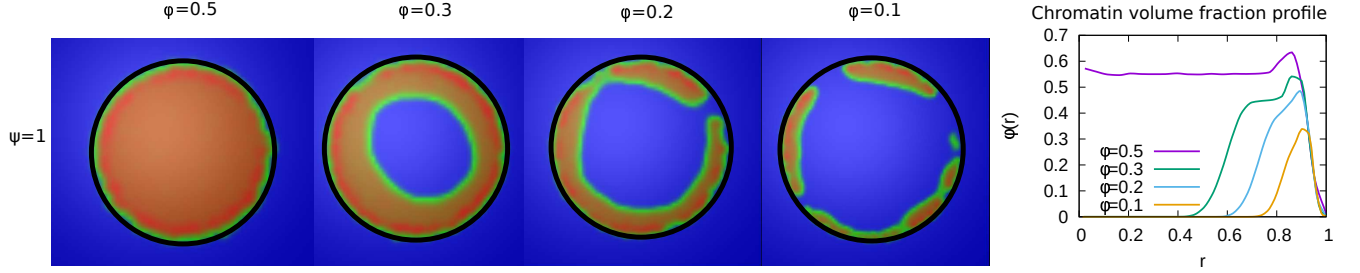}
\caption{{\it Left panel:} Chromatin concentrations are shown for different 
volume fraction of chromatin, $\phi$. {\it Right panel:} The \tb{local volume 
fraction} profiles show peripheral organization for  $\phi=0.3$ and $\phi=0.1$ 
but not for $\phi=0.5$.  This demonstrates the transition from  peripheral to 
conventional chromatin organization as the nucleus is dehydrated and $\phi$ is 
increased.  
\label{fig:04}
 } 
\end{center}
\end{figure}

{\bf Effect of hydration (variation of chromatin volume fraction) of the nucleus:} 
In the experiments, the volume of live {\it Drosophila} larva nuclei had an average value of $1183 um^3$, while fixed nuclei had an average volume of $381 um^3$, with corresponding changes in the chromatin organization from peripheral to conventional. This motivated our simulation study of the effect of changes in the relative fractions of the chromatin on chromatin organization, by changing the parameter $\phi$. For relatively high chromatin volume fractions ($\phi=0.5$), chromatin fills the entire volume of nucleus and shows conventional organization; there is no phase separation of chromatin and aqueous phase due to the overabundance of chromatin in the system (left panel of Fig.~\ref{fig:04}). As we decrease the volume fraction of chromatin, the organization shows a transition from conventional to one that is more peripheral. In this case, there is sufficient aqueous phase so that phase separation is possible.

\begin{figure} [H] 
\begin{center}
\includegraphics[scale=0.7]{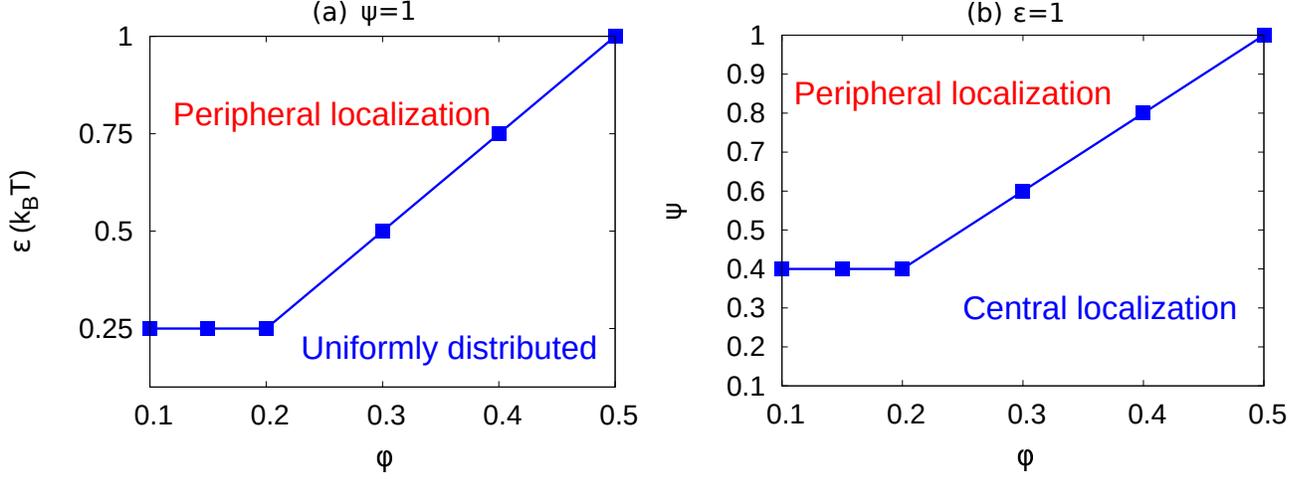}
\caption{(a) For a fixed value of $\psi=1$ (maximally bonded LAD), we 
calculated the \tb{local volume fraction} obtained from simulations with 
different pairs of the chromatin volume fraction and self-attraction 
$(\phi,{\epsilon})$ where $0.1\leq\phi\leq 0.5$ and $0.25\leq{ \epsilon}\leq 
1$. For each pair of $(\phi,{\epsilon})$, we used the plots of the \tb{local 
volume fraction}  to determine the chromatin organization mode. In the graph, 
the blue line shows the transition between conventional (uniformly distributed) 
chromatin organization, and  peripheral organization. (b) For a fixed value of 
the chromatin self-attraction $\epsilon=1$ we varied the chromatin volume 
fraction and fraction of LAD that can bond to the lamina $(\phi,\psi)$ where 
$0.1\leq\phi\leq 0.5$ and $0.1\leq\psi\leq 1$. The blue line shows the 
transition from central (and wetting drop) to peripheral chromatin organization.
\label{fig:05}
 } 
\end{center}
\end{figure}

{\bf Transitions from peripheral to uniform to central localization of 
chromatin:} In the results shown above, we see how chromatin organization 
changes when we change $\phi,\psi$ and $\epsilon$. Here, we summarize the 
transitions of chromatin organization from peripheral to central or from 
peripheral to uniform via a state diagram. We did many simulations for 
different values of the parameters $(\phi,\psi,\epsilon)$ and calculated 
\tb{local volume fraction} profiles for these ranges. From the plots of 
\tb{local volume fraction}, we classified the chromatin organization as 
peripheral, central (which can include the wetting droplet configurations 
discussed in the SI text and Fig.S7) or conventional. In Fig.~\ref{fig:05}, we present the 
state diagram showing the transitions in chromatin organization by variations 
of the chromatin volume fraction, fraction of LAD that can bond to the lamin, 
and the intra-chromatin attraction strength $(\phi,\psi,{\epsilon})$. These 
results show how this minimal set of coarse-grained, simulation parameters 
$(\phi,\psi,{\epsilon})$ already determines a rich variety of  chromatin 
organization in the nucleus.

\section*{Discussion} \tr{We have shown here that the observed transitions in nuclear-scale chromatin organization in eukaryotic cells can be understood from the physics of a simple polymer model. The comparison of the theoretical and experimental trends indicates that these transitions are due to competition of polymer entropy, self-attraction of the chromatin and the chromatin-lamin attraction. By varying the values of the three generic parameters ($\phi, \epsilon, \psi$) that respectively control these effects, our simulations demonstrated the transitions between peripheral, central and conventional organization of chromatin as seen in the experiments. }

\tr{For example, peripheral organization of chromatin was observed in our simulations when:  (i) The radius of confinement within the nucleus is  greater than radius of gyration of chromatin ($R_c>R_g$), which is relevant for relatively small volume fractions of chromatin ($\phi=0.3$). (ii) Chromatin must be self-attractive for its localization on the nuclear periphery, which occurs for significant chromatin self-attraction ($\epsilon=1$);. (iii) Chromatin-lamina interactions are strong, as in the simulations for the case where all (or most) of the LADs can bond to the lamina ($\psi=1$) (see Fig.~\ref{fig:01}). }

\tr{Starting with these parameter values that resulted in peripheral organization, we then varied $\phi,\psi,\epsilon$ to determine from our simulations, the transitions from peripheral to central to conventional, by varying one parameter at a time. We found  transitions from peripheral to conventional when the self-attraction and volume fractions were decreased: $0.25\leq\epsilon\leq1$ and $0.1\leq\phi\leq0.5$ (see Fig.~\ref{fig:02} and Fig.~\ref{fig:04}). We also found transitions from peripheral to central by reducing the chromatin-lamina interactions ($0\leq\psi\leq1$) (see Fig.~\ref{fig:03}). We plotted the  state diagrams for these transitions in Fig.~\ref{fig:05}. These qualitatively track the observed experimental trends  for example, peripheral to conventional chromatin organization when the cell is dehydrated so that the chromatin volume fraction is increased. In the SI (Fig.~S8) we discuss in more detail the comparison of the simulations and the experimental images that are published elsewhere~\cite{Amiad-pavlov2020}.}

\begin{figure} [H] 
\begin{center}
\includegraphics[scale=0.35]{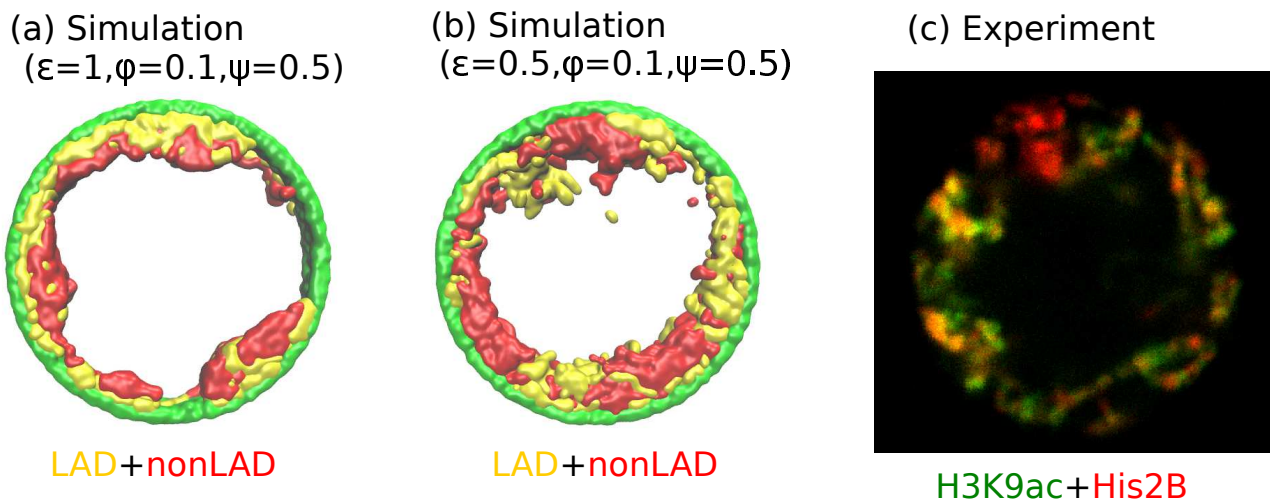}
\caption{(a) Simulation snapshot for parameter values $\psi=1$, $\phi=0.1$, 
and $\epsilon=0.5$, shows  peripheral organization with LAD near lamina and 
non-LADs separated from LADs in the radial direction. (b) Simulation snapshot 
for parameter values 
$\psi=0.5$, $\phi=0.1$, and $\epsilon=0.5$, shows peripheral organization with 
alternating LAD and non-LAD regions in the angular direction at the nuclear 
periphery. (c) Experimentally labeled H3K9ac (euchromatin/most of nonLAD/green) 
and His2B (chromatin/LAD and non-LAD/red) in muscle nuclei  of \tr{intact} {\it 
Drosophila}  larvae, shows heterochromatin (associated with LAD) by dark red 
color and euchromatin (associated with non-LAD) by merging  the red and green 
colors in peripheral organization. Both the experiments and simulations show an 
angular 
distribution of LAD and non-LAD as opposed to a radial distribution.} 
\label{fig:06}
 
\end{center}
\end{figure}

\tbb{An interesting experimental feature observed in the peripheral organization of chromatin, is that euchromatin (active chromatin, typically associated with  non-LAD) and heterochromatin (inactive chromatin, typically associated with LAD) regions do not separate in the radial directions, but are separated in the perpendicular (angular) direction (see Fig.~\ref{fig:06}(c)). Our simulations show that this can be the case even if the self-attraction of LAD and non-LAD are the same; this corresponds to the limit of small differences in heterochromatin/euchromatin interactions relative to their interactions with the aqueous phase and with the lamina. In this case, the angular separation originates in the stronger binding of the LAD domains to the lamina.  We observed the angular separation of LAD and non-LAD in our simulations, when we allowed both the LAD and the non-LAD to have LJ attractions to the lamina (arising from physical interactions such as van der Waals) but allow only the LAD to additionally forms biochemical bonds with the lamina (Biophysically, this corresponds to bonds formed via proteins such as BAF). The angular alternation of LAD and non-LAD on the nuclear pheriphery occurs when only a fraction of the LAD can bind to the lamina (e.g., due to the limited binding proteins) corresponding to $\psi=0.5$,  and for $\phi=0.1$, and $\epsilon=0.5$, (see Fig.~\ref{fig:06}(b)). These values of our simulation parameters also correspond to the experimental conditions. Small volume fractions of chromatin ($\phi=0.1$) correspond experimentally to large (hydrated) {\it Drosophila} nuclei~\cite{Amiad-pavlov2020}. We note that for these simulations a somewhat smaller self-attraction,  $\epsilon=0.5$, shows the angular separation of LAD and non-LAD chromatin as seen in the experiments, whereas for $\epsilon=1$, we find homogeneous peripheral distribution with mostly LAD domains near the lamina, similar to Fig.~\ref{fig:06}(a). }

\trr{Although the qualitative trends in the simulations and experiments are indeed similar, it would be interesting for new experiments to more quantitatively test our state diagram and the predicted transitions in Fig.~\ref{fig:05}}. For example, {\it in vitro} experiments could control the amount of dehydration (changing the chromosome fraction $\phi$ with fixed chromosome mass) by spreading cells on surfaces~\cite{Guo2017,Adar5604}.

\tr{To focus on the essential physics of chromatin self-attraction vs. chromatin-lamin interactions, we have simplified the chromatin self-attractions to be the same for the euchromatin (EC) and heterochromatin (HC) domains.   This is appropriate when the chromatin-aqueous phase interaction and the lamin-chromatin interaction are larger than the differences in the interactions of euchromatin and heterochromatin. Our model is the minimal one required to understand how the chromatin separates from the aqueous phase in peripheral and central organization. Although, previous models explicitly accounted for different interactions of euchromatin and heterochromatin domains, they did not consider the various types of nuclear-scale chromatin/aqueous phase/laminar organization and typically focused only on  conventional chromatin organization. Extending our approach  to also include different EC and HC interactions can predict HC/EC micro-phase separation (termed, A/B compartments) at smaller distances than the nuclear scale that the experiments~\cite{Amiad-pavlov2020} and our simulations have considered.  }

{\it Acknowledgements:} SAS is grateful for the Kretner-Katz and Perlman family foundations, the US-Israel Binational Science Foundation, and the Volkswagen Foundation for their support. We thanks D. Deviri, Omar Adame-Arana, Phil Pincus, and Michael Rubinstein for useful discussions.
~\\

{\it Author contributions:} GB and SS performed and analyzed the simulations. DL, DAP, and TV performed and analyzed the experiments.  The paper was written by GB and SS with substantial input by DL, DAP, and TV.
\newpage

\bibliography{Black_hole_work}

\begin{thebibliography}{41}
\expandafter\ifx\csname natexlab\endcsname\relax\def\natexlab#1{#1}\fi
\expandafter\ifx\csname bibnamefont\endcsname\relax
  \def\bibnamefont#1{#1}\fi
\expandafter\ifx\csname bibfnamefont\endcsname\relax
  \def\bibfnamefont#1{#1}\fi
\expandafter\ifx\csname citenamefont\endcsname\relax
  \def\citenamefont#1{#1}\fi
\expandafter\ifx\csname url\endcsname\relax
  \def\url#1{\texttt{#1}}\fi
\expandafter\ifx\csname urlprefix\endcsname\relax\def\urlprefix{URL }\fi
\providecommand{\bibinfo}[2]{#2}
\providecommand{\eprint}[2][]{\url{#2}}

\bibitem[{\citenamefont{Cooper and Hausman}(2000)}]{cooper2000molecular}
\bibinfo{author}{\bibfnamefont{G.~M.} \bibnamefont{Cooper}} \bibnamefont{and}
  \bibinfo{author}{\bibfnamefont{R.}~\bibnamefont{Hausman}},
  \bibinfo{journal}{The Cell. 2nd ed. Sunderland, MA: Sinauer Associates}
  (\bibinfo{year}{2000}).

\bibitem[{\citenamefont{Phillips et~al.}(2012)\citenamefont{Phillips, Kondev,
  Theriot, and Garcia}}]{phillips2012physical}
\bibinfo{author}{\bibfnamefont{R.}~\bibnamefont{Phillips}},
  \bibinfo{author}{\bibfnamefont{J.}~\bibnamefont{Kondev}},
  \bibinfo{author}{\bibfnamefont{J.}~\bibnamefont{Theriot}}, \bibnamefont{and}
  \bibinfo{author}{\bibfnamefont{H.}~\bibnamefont{Garcia}},
  \emph{\bibinfo{title}{Physical biology of the cell}}
  (\bibinfo{publisher}{Garland Science}, \bibinfo{year}{2012}).

\bibitem[{\citenamefont{Rosa and Shaw}(2013)}]{rosa2013insights}
\bibinfo{author}{\bibfnamefont{S.}~\bibnamefont{Rosa}} \bibnamefont{and}
  \bibinfo{author}{\bibfnamefont{P.}~\bibnamefont{Shaw}},
  \bibinfo{journal}{Biology} \textbf{\bibinfo{volume}{2}},
  \bibinfo{pages}{1378} (\bibinfo{year}{2013}).

\bibitem[{\citenamefont{Rubinstein et~al.}(2003)\citenamefont{Rubinstein, Colby
  et~al.}}]{rubinstein2003polymer}
\bibinfo{author}{\bibfnamefont{M.}~\bibnamefont{Rubinstein}},
  \bibinfo{author}{\bibfnamefont{R.~H.} \bibnamefont{Colby}},
  \bibnamefont{et~al.}, \emph{\bibinfo{title}{Polymer physics}},
  vol.~\bibinfo{volume}{23} (\bibinfo{publisher}{Oxford university press New
  York}, \bibinfo{year}{2003}).

\bibitem[{\citenamefont{Erdel and Rippe}(2018)}]{erdel2018formation}
\bibinfo{author}{\bibfnamefont{F.}~\bibnamefont{Erdel}} \bibnamefont{and}
  \bibinfo{author}{\bibfnamefont{K.}~\bibnamefont{Rippe}},
  \bibinfo{journal}{Biophysical journal} \textbf{\bibinfo{volume}{114}},
  \bibinfo{pages}{2262} (\bibinfo{year}{2018}).

\bibitem[{\citenamefont{Narlikar}(2020)}]{narlikar2020phase}
\bibinfo{author}{\bibfnamefont{G.~J.} \bibnamefont{Narlikar}},
  \bibinfo{journal}{Journal of Biosciences} \textbf{\bibinfo{volume}{45}},
  \bibinfo{pages}{5} (\bibinfo{year}{2020}).

\bibitem[{\citenamefont{Strom et~al.}(2017)\citenamefont{Strom, Emelyanov, Mir,
  Fyodorov, Darzacq, and Karpen}}]{Strom2017}
\bibinfo{author}{\bibfnamefont{A.~R.} \bibnamefont{Strom}},
  \bibinfo{author}{\bibfnamefont{A.~V.} \bibnamefont{Emelyanov}},
  \bibinfo{author}{\bibfnamefont{M.}~\bibnamefont{Mir}},
  \bibinfo{author}{\bibfnamefont{D.~V.} \bibnamefont{Fyodorov}},
  \bibinfo{author}{\bibfnamefont{X.}~\bibnamefont{Darzacq}}, \bibnamefont{and}
  \bibinfo{author}{\bibfnamefont{G.~H.} \bibnamefont{Karpen}},
  \bibinfo{journal}{Nature} \textbf{\bibinfo{volume}{547}},
  \bibinfo{pages}{241} (\bibinfo{year}{2017}).

\bibitem[{\citenamefont{Larson et~al.}(2017)\citenamefont{Larson, Elnatan,
  Keenen, Trnka, Johnston, Burlingame, Agard, Redding, and
  Narlikar}}]{Larson2017}
\bibinfo{author}{\bibfnamefont{A.~G.} \bibnamefont{Larson}},
  \bibinfo{author}{\bibfnamefont{D.}~\bibnamefont{Elnatan}},
  \bibinfo{author}{\bibfnamefont{M.~M.} \bibnamefont{Keenen}},
  \bibinfo{author}{\bibfnamefont{M.~J.} \bibnamefont{Trnka}},
  \bibinfo{author}{\bibfnamefont{J.~B.} \bibnamefont{Johnston}},
  \bibinfo{author}{\bibfnamefont{A.~L.} \bibnamefont{Burlingame}},
  \bibinfo{author}{\bibfnamefont{D.~A.} \bibnamefont{Agard}},
  \bibinfo{author}{\bibfnamefont{S.}~\bibnamefont{Redding}}, \bibnamefont{and}
  \bibinfo{author}{\bibfnamefont{G.~J.} \bibnamefont{Narlikar}},
  \bibinfo{journal}{Nature} \textbf{\bibinfo{volume}{547}},
  \bibinfo{pages}{236} (\bibinfo{year}{2017}).

\bibitem[{\citenamefont{Lieberman-aiden
  et~al.}(2009)\citenamefont{Lieberman-aiden, Berkum, Williams, Imakaev,
  Ragoczy, Telling, Amit, Lajoie, Sabo, Dorschner
  et~al.}}]{Lieberman-aiden2009}
\bibinfo{author}{\bibfnamefont{E.}~\bibnamefont{Lieberman-aiden}},
  \bibinfo{author}{\bibfnamefont{N.~L.~V.} \bibnamefont{Berkum}},
  \bibinfo{author}{\bibfnamefont{L.}~\bibnamefont{Williams}},
  \bibinfo{author}{\bibfnamefont{M.}~\bibnamefont{Imakaev}},
  \bibinfo{author}{\bibfnamefont{T.}~\bibnamefont{Ragoczy}},
  \bibinfo{author}{\bibfnamefont{A.}~\bibnamefont{Telling}},
  \bibinfo{author}{\bibfnamefont{I.}~\bibnamefont{Amit}},
  \bibinfo{author}{\bibfnamefont{B.~R.} \bibnamefont{Lajoie}},
  \bibinfo{author}{\bibfnamefont{P.~J.} \bibnamefont{Sabo}},
  \bibinfo{author}{\bibfnamefont{M.~O.} \bibnamefont{Dorschner}},
  \bibnamefont{et~al.}, \bibinfo{journal}{Science}
  \textbf{\bibinfo{volume}{33292}}, \bibinfo{pages}{289}
  (\bibinfo{year}{2009}).

\bibitem[{\citenamefont{Cremer et~al.}(2015)\citenamefont{Cremer, Cremer,
  H{\"u}bner, Strickfaden, Smeets, Popken, Sterr, Markaki, Rippe, and
  Cremer}}]{cremer20154d}
\bibinfo{author}{\bibfnamefont{T.}~\bibnamefont{Cremer}},
  \bibinfo{author}{\bibfnamefont{M.}~\bibnamefont{Cremer}},
  \bibinfo{author}{\bibfnamefont{B.}~\bibnamefont{H{\"u}bner}},
  \bibinfo{author}{\bibfnamefont{H.}~\bibnamefont{Strickfaden}},
  \bibinfo{author}{\bibfnamefont{D.}~\bibnamefont{Smeets}},
  \bibinfo{author}{\bibfnamefont{J.}~\bibnamefont{Popken}},
  \bibinfo{author}{\bibfnamefont{M.}~\bibnamefont{Sterr}},
  \bibinfo{author}{\bibfnamefont{Y.}~\bibnamefont{Markaki}},
  \bibinfo{author}{\bibfnamefont{K.}~\bibnamefont{Rippe}}, \bibnamefont{and}
  \bibinfo{author}{\bibfnamefont{C.}~\bibnamefont{Cremer}},
  \bibinfo{journal}{FEBS letters} \textbf{\bibinfo{volume}{589}},
  \bibinfo{pages}{2931} (\bibinfo{year}{2015}).

\bibitem[{\citenamefont{Popken et~al.}(2014)\citenamefont{Popken, Brero,
  Koehler, Schmid, Strauss, Wuensch, Guengoer, Graf, Krebs, Blum
  et~al.}}]{Popken2014}
\bibinfo{author}{\bibfnamefont{J.}~\bibnamefont{Popken}},
  \bibinfo{author}{\bibfnamefont{A.}~\bibnamefont{Brero}},
  \bibinfo{author}{\bibfnamefont{D.}~\bibnamefont{Koehler}},
  \bibinfo{author}{\bibfnamefont{V.~J.} \bibnamefont{Schmid}},
  \bibinfo{author}{\bibfnamefont{A.}~\bibnamefont{Strauss}},
  \bibinfo{author}{\bibfnamefont{A.}~\bibnamefont{Wuensch}},
  \bibinfo{author}{\bibfnamefont{T.}~\bibnamefont{Guengoer}},
  \bibinfo{author}{\bibfnamefont{A.}~\bibnamefont{Graf}},
  \bibinfo{author}{\bibfnamefont{S.}~\bibnamefont{Krebs}},
  \bibinfo{author}{\bibfnamefont{H.}~\bibnamefont{Blum}}, \bibnamefont{et~al.},
  \bibinfo{journal}{Nucleus} \textbf{\bibinfo{volume}{5}}, \bibinfo{pages}{555}
  (\bibinfo{year}{2014}).

\bibitem[{\citenamefont{Amiad-Pavlov et~al.}(2020)\citenamefont{Amiad-Pavlov,
  Lorber, Bajpai, Safran, and Volk}}]{Amiad-pavlov2020}
\bibinfo{author}{\bibfnamefont{D.}~\bibnamefont{Amiad-Pavlov}},
  \bibinfo{author}{\bibfnamefont{D.}~\bibnamefont{Lorber}},
  \bibinfo{author}{\bibfnamefont{G.}~\bibnamefont{Bajpai}},
  \bibinfo{author}{\bibfnamefont{S.}~\bibnamefont{Safran}}, \bibnamefont{and}
  \bibinfo{author}{\bibfnamefont{T.}~\bibnamefont{Volk}},
  \bibinfo{journal}{bioRxiv}  (\bibinfo{year}{2020}).

\bibitem[{\citenamefont{Ulianov et~al.}(2019)\citenamefont{Ulianov, Doronin,
  Khrameeva, Kos, Luzhin, Starikov, Galitsyna, Nenasheva, Ilyin, Flyamer
  et~al.}}]{ulianov2019nuclear}
\bibinfo{author}{\bibfnamefont{S.~V.} \bibnamefont{Ulianov}},
  \bibinfo{author}{\bibfnamefont{S.~A.} \bibnamefont{Doronin}},
  \bibinfo{author}{\bibfnamefont{E.~E.} \bibnamefont{Khrameeva}},
  \bibinfo{author}{\bibfnamefont{P.~I.} \bibnamefont{Kos}},
  \bibinfo{author}{\bibfnamefont{A.~V.} \bibnamefont{Luzhin}},
  \bibinfo{author}{\bibfnamefont{S.~S.} \bibnamefont{Starikov}},
  \bibinfo{author}{\bibfnamefont{A.~A.} \bibnamefont{Galitsyna}},
  \bibinfo{author}{\bibfnamefont{V.~V.} \bibnamefont{Nenasheva}},
  \bibinfo{author}{\bibfnamefont{A.~A.} \bibnamefont{Ilyin}},
  \bibinfo{author}{\bibfnamefont{I.~M.} \bibnamefont{Flyamer}},
  \bibnamefont{et~al.}, \bibinfo{journal}{Nature communications}
  \textbf{\bibinfo{volume}{10}}, \bibinfo{pages}{1} (\bibinfo{year}{2019}).

\bibitem[{\citenamefont{Kind et~al.}(2013)\citenamefont{Kind, Pagie,
  Ortabozkoyun, Boyle, {De Vries}, Janssen, Amendola, Nolen, Bickmore, and {Van
  Steensel}}}]{Kind2013}
\bibinfo{author}{\bibfnamefont{J.}~\bibnamefont{Kind}},
  \bibinfo{author}{\bibfnamefont{L.}~\bibnamefont{Pagie}},
  \bibinfo{author}{\bibfnamefont{H.}~\bibnamefont{Ortabozkoyun}},
  \bibinfo{author}{\bibfnamefont{S.}~\bibnamefont{Boyle}},
  \bibinfo{author}{\bibfnamefont{S.~S.} \bibnamefont{{De Vries}}},
  \bibinfo{author}{\bibfnamefont{H.}~\bibnamefont{Janssen}},
  \bibinfo{author}{\bibfnamefont{M.}~\bibnamefont{Amendola}},
  \bibinfo{author}{\bibfnamefont{L.~D.} \bibnamefont{Nolen}},
  \bibinfo{author}{\bibfnamefont{W.~A.} \bibnamefont{Bickmore}},
  \bibnamefont{and} \bibinfo{author}{\bibfnamefont{B.}~\bibnamefont{{Van
  Steensel}}}, \bibinfo{journal}{Cell} \textbf{\bibinfo{volume}{153}},
  \bibinfo{pages}{178} (\bibinfo{year}{2013}).

\bibitem[{\citenamefont{Y{\'{a}}{\~{n}}ez-Cuna and van
  Steensel}(2017)}]{Yanez-Cuna2017a}
\bibinfo{author}{\bibfnamefont{J.~O.} \bibnamefont{Y{\'{a}}{\~{n}}ez-Cuna}}
  \bibnamefont{and} \bibinfo{author}{\bibfnamefont{B.}~\bibnamefont{van
  Steensel}}, \bibinfo{journal}{Current Opinion in Genetics and Development}
  \textbf{\bibinfo{volume}{43}}, \bibinfo{pages}{67} (\bibinfo{year}{2017}).

\bibitem[{\citenamefont{Moir et~al.}(2000)\citenamefont{Moir, Yoon, Khuon, and
  Goldman}}]{moir2000nuclear}
\bibinfo{author}{\bibfnamefont{R.~D.} \bibnamefont{Moir}},
  \bibinfo{author}{\bibfnamefont{M.}~\bibnamefont{Yoon}},
  \bibinfo{author}{\bibfnamefont{S.}~\bibnamefont{Khuon}}, \bibnamefont{and}
  \bibinfo{author}{\bibfnamefont{R.~D.} \bibnamefont{Goldman}},
  \bibinfo{journal}{The Journal of cell biology}
  \textbf{\bibinfo{volume}{151}}, \bibinfo{pages}{1155} (\bibinfo{year}{2000}).

\bibitem[{\citenamefont{Riemer et~al.}(1995)\citenamefont{Riemer, Stuurman,
  Berrios, Hunter, Fisher, and Weber}}]{riemer1995expression}
\bibinfo{author}{\bibfnamefont{D.}~\bibnamefont{Riemer}},
  \bibinfo{author}{\bibfnamefont{N.}~\bibnamefont{Stuurman}},
  \bibinfo{author}{\bibfnamefont{M.}~\bibnamefont{Berrios}},
  \bibinfo{author}{\bibfnamefont{C.}~\bibnamefont{Hunter}},
  \bibinfo{author}{\bibfnamefont{P.~A.} \bibnamefont{Fisher}},
  \bibnamefont{and} \bibinfo{author}{\bibfnamefont{K.}~\bibnamefont{Weber}},
  \bibinfo{journal}{Journal of cell science} \textbf{\bibinfo{volume}{108}},
  \bibinfo{pages}{3189} (\bibinfo{year}{1995}).

\bibitem[{\citenamefont{Schulze et~al.}(2009)\citenamefont{Schulze,
  Curio-Penny, Speese, Dialynas, Cryderman, McDonough, Nalbant, Petersen,
  Budnik, Geyer et~al.}}]{schulze2009comparative}
\bibinfo{author}{\bibfnamefont{S.~R.} \bibnamefont{Schulze}},
  \bibinfo{author}{\bibfnamefont{B.}~\bibnamefont{Curio-Penny}},
  \bibinfo{author}{\bibfnamefont{S.}~\bibnamefont{Speese}},
  \bibinfo{author}{\bibfnamefont{G.}~\bibnamefont{Dialynas}},
  \bibinfo{author}{\bibfnamefont{D.~E.} \bibnamefont{Cryderman}},
  \bibinfo{author}{\bibfnamefont{C.~W.} \bibnamefont{McDonough}},
  \bibinfo{author}{\bibfnamefont{D.}~\bibnamefont{Nalbant}},
  \bibinfo{author}{\bibfnamefont{M.}~\bibnamefont{Petersen}},
  \bibinfo{author}{\bibfnamefont{V.}~\bibnamefont{Budnik}},
  \bibinfo{author}{\bibfnamefont{P.~K.} \bibnamefont{Geyer}},
  \bibnamefont{et~al.}, \bibinfo{journal}{PLoS One}
  \textbf{\bibinfo{volume}{4}}, \bibinfo{pages}{e7564} (\bibinfo{year}{2009}).

\bibitem[{\citenamefont{Briand and Collas}(2020)}]{Briand2020}
\bibinfo{author}{\bibfnamefont{N.}~\bibnamefont{Briand}} \bibnamefont{and}
  \bibinfo{author}{\bibfnamefont{P.}~\bibnamefont{Collas}},
  \bibinfo{journal}{Genome Biology} \textbf{\bibinfo{volume}{21}},
  \bibinfo{pages}{1} (\bibinfo{year}{2020}).

\bibitem[{\citenamefont{Naetar et~al.}(2017)\citenamefont{Naetar, Ferraioli,
  and Foisner}}]{naetar2017lamins}
\bibinfo{author}{\bibfnamefont{N.}~\bibnamefont{Naetar}},
  \bibinfo{author}{\bibfnamefont{S.}~\bibnamefont{Ferraioli}},
  \bibnamefont{and} \bibinfo{author}{\bibfnamefont{R.}~\bibnamefont{Foisner}},
  \bibinfo{journal}{Journal of cell science} \textbf{\bibinfo{volume}{130}},
  \bibinfo{pages}{2087} (\bibinfo{year}{2017}).

\bibitem[{\citenamefont{van Steensel and Belmont}(2017)}]{VanSteensel2017}
\bibinfo{author}{\bibfnamefont{B.}~\bibnamefont{van Steensel}}
  \bibnamefont{and} \bibinfo{author}{\bibfnamefont{A.~S.}
  \bibnamefont{Belmont}}, \bibinfo{journal}{Cell}
  \textbf{\bibinfo{volume}{169}}, \bibinfo{pages}{780} (\bibinfo{year}{2017}).

\bibitem[{\citenamefont{Solovei et~al.}(2013)\citenamefont{Solovei, Wang,
  Thanisch, Schmidt, Krebs, Zwerger, Cohen, Devys, Foisner, Peichl
  et~al.}}]{Solovei2013}
\bibinfo{author}{\bibfnamefont{I.}~\bibnamefont{Solovei}},
  \bibinfo{author}{\bibfnamefont{A.~S.} \bibnamefont{Wang}},
  \bibinfo{author}{\bibfnamefont{K.}~\bibnamefont{Thanisch}},
  \bibinfo{author}{\bibfnamefont{C.~S.} \bibnamefont{Schmidt}},
  \bibinfo{author}{\bibfnamefont{S.}~\bibnamefont{Krebs}},
  \bibinfo{author}{\bibfnamefont{M.}~\bibnamefont{Zwerger}},
  \bibinfo{author}{\bibfnamefont{T.~V.} \bibnamefont{Cohen}},
  \bibinfo{author}{\bibfnamefont{D.}~\bibnamefont{Devys}},
  \bibinfo{author}{\bibfnamefont{R.}~\bibnamefont{Foisner}},
  \bibinfo{author}{\bibfnamefont{L.}~\bibnamefont{Peichl}},
  \bibnamefont{et~al.}, \bibinfo{journal}{Cell} \textbf{\bibinfo{volume}{152}},
  \bibinfo{pages}{584} (\bibinfo{year}{2013}).

\bibitem[{\citenamefont{Wagner et~al.}(2004)\citenamefont{Wagner, Weber, Seitz,
  and Krohne}}]{wagner2004lamin}
\bibinfo{author}{\bibfnamefont{N.}~\bibnamefont{Wagner}},
  \bibinfo{author}{\bibfnamefont{D.}~\bibnamefont{Weber}},
  \bibinfo{author}{\bibfnamefont{S.}~\bibnamefont{Seitz}}, \bibnamefont{and}
  \bibinfo{author}{\bibfnamefont{G.}~\bibnamefont{Krohne}},
  \bibinfo{journal}{Journal of cell science} \textbf{\bibinfo{volume}{117}},
  \bibinfo{pages}{2015} (\bibinfo{year}{2004}).

\bibitem[{\citenamefont{Chiang et~al.}(2019)\citenamefont{Chiang, Michieletto,
  Brackley, Rattanavirotkul, Mohammed, Marenduzzo, and Chandra}}]{Chiang2019}
\bibinfo{author}{\bibfnamefont{M.}~\bibnamefont{Chiang}},
  \bibinfo{author}{\bibfnamefont{D.}~\bibnamefont{Michieletto}},
  \bibinfo{author}{\bibfnamefont{C.~A.} \bibnamefont{Brackley}},
  \bibinfo{author}{\bibfnamefont{N.}~\bibnamefont{Rattanavirotkul}},
  \bibinfo{author}{\bibfnamefont{H.}~\bibnamefont{Mohammed}},
  \bibinfo{author}{\bibfnamefont{D.}~\bibnamefont{Marenduzzo}},
  \bibnamefont{and} \bibinfo{author}{\bibfnamefont{T.}~\bibnamefont{Chandra}},
  \bibinfo{journal}{Cell Reports} \textbf{\bibinfo{volume}{28}},
  \bibinfo{pages}{3212} (\bibinfo{year}{2019}).

\bibitem[{\citenamefont{Maji et~al.}(2020)\citenamefont{Maji, Ahmed, Roy,
  Chakrabarti, and Mitra}}]{Maji2020}
\bibinfo{author}{\bibfnamefont{A.}~\bibnamefont{Maji}},
  \bibinfo{author}{\bibfnamefont{J.~A.} \bibnamefont{Ahmed}},
  \bibinfo{author}{\bibfnamefont{S.}~\bibnamefont{Roy}},
  \bibinfo{author}{\bibfnamefont{B.}~\bibnamefont{Chakrabarti}},
  \bibnamefont{and} \bibinfo{author}{\bibfnamefont{M.~K.} \bibnamefont{Mitra}},
  \bibinfo{journal}{Biophysical Journal} \textbf{\bibinfo{volume}{118}},
  \bibinfo{pages}{3041} (\bibinfo{year}{2020}).

\bibitem[{\citenamefont{Gibson et~al.}(2019)\citenamefont{Gibson, Doolittle,
  Schneider, Jensen, Gamarra, Henry, Gerlich, Redding, and Rosen}}]{Rosen2019}
\bibinfo{author}{\bibfnamefont{B.~A.} \bibnamefont{Gibson}},
  \bibinfo{author}{\bibfnamefont{L.~K.} \bibnamefont{Doolittle}},
  \bibinfo{author}{\bibfnamefont{M.~W.~G.} \bibnamefont{Schneider}},
  \bibinfo{author}{\bibfnamefont{L.~E.} \bibnamefont{Jensen}},
  \bibinfo{author}{\bibfnamefont{N.}~\bibnamefont{Gamarra}},
  \bibinfo{author}{\bibfnamefont{L.}~\bibnamefont{Henry}},
  \bibinfo{author}{\bibfnamefont{D.~W.} \bibnamefont{Gerlich}},
  \bibinfo{author}{\bibfnamefont{S.}~\bibnamefont{Redding}}, \bibnamefont{and}
  \bibinfo{author}{\bibfnamefont{M.~K.} \bibnamefont{Rosen}},
  \bibinfo{journal}{Cell} \textbf{\bibinfo{volume}{179}}, \bibinfo{pages}{470 }
  (\bibinfo{year}{2019}).

\bibitem[{\citenamefont{Bajpai and Padinhateeri}(2019)}]{Bajpai2019}
\bibinfo{author}{\bibfnamefont{G.}~\bibnamefont{Bajpai}} \bibnamefont{and}
  \bibinfo{author}{\bibfnamefont{R.}~\bibnamefont{Padinhateeri}},
  \bibinfo{journal}{Biophysical Journal} \textbf{\bibinfo{volume}{118}},
  \bibinfo{pages}{207} (\bibinfo{year}{2019}).

\bibitem[{\citenamefont{Hancock}(2007)}]{hancock2007packing}
\bibinfo{author}{\bibfnamefont{R.}~\bibnamefont{Hancock}}, in
  \emph{\bibinfo{booktitle}{Seminars in cell \& developmental biology}}
  (\bibinfo{organization}{Elsevier}, \bibinfo{year}{2007}),
  vol.~\bibinfo{volume}{18}, pp. \bibinfo{pages}{668--675}.

\bibitem[{\citenamefont{Bronshtein et~al.}(2016)\citenamefont{Bronshtein,
  Kanter, Kepten, Lindner, Berezin, Shav-Tal, and Garini}}]{Bronshtein2016}
\bibinfo{author}{\bibfnamefont{I.}~\bibnamefont{Bronshtein}},
  \bibinfo{author}{\bibfnamefont{I.}~\bibnamefont{Kanter}},
  \bibinfo{author}{\bibfnamefont{E.}~\bibnamefont{Kepten}},
  \bibinfo{author}{\bibfnamefont{M.}~\bibnamefont{Lindner}},
  \bibinfo{author}{\bibfnamefont{S.}~\bibnamefont{Berezin}},
  \bibinfo{author}{\bibfnamefont{Y.}~\bibnamefont{Shav-Tal}}, \bibnamefont{and}
  \bibinfo{author}{\bibfnamefont{Y.}~\bibnamefont{Garini}},
  \bibinfo{journal}{Nucleus} \textbf{\bibinfo{volume}{7}}, \bibinfo{pages}{27}
  (\bibinfo{year}{2016}).

\bibitem[{\citenamefont{Bronshtein et~al.}(2015)\citenamefont{Bronshtein,
  Kepten, Kanter, Berezin, Lindner, Redwood, Mai, Gonzalo, Foisner, Shav-Tal
  et~al.}}]{Bronshtein2015}
\bibinfo{author}{\bibfnamefont{I.}~\bibnamefont{Bronshtein}},
  \bibinfo{author}{\bibfnamefont{E.}~\bibnamefont{Kepten}},
  \bibinfo{author}{\bibfnamefont{I.}~\bibnamefont{Kanter}},
  \bibinfo{author}{\bibfnamefont{S.}~\bibnamefont{Berezin}},
  \bibinfo{author}{\bibfnamefont{M.}~\bibnamefont{Lindner}},
  \bibinfo{author}{\bibfnamefont{A.~B.} \bibnamefont{Redwood}},
  \bibinfo{author}{\bibfnamefont{S.}~\bibnamefont{Mai}},
  \bibinfo{author}{\bibfnamefont{S.}~\bibnamefont{Gonzalo}},
  \bibinfo{author}{\bibfnamefont{R.}~\bibnamefont{Foisner}},
  \bibinfo{author}{\bibfnamefont{Y.}~\bibnamefont{Shav-Tal}},
  \bibnamefont{et~al.}, \bibinfo{journal}{Nature Communications}
  \textbf{\bibinfo{volume}{6}}, \bibinfo{pages}{1} (\bibinfo{year}{2015}).

\bibitem[{\citenamefont{Stephens et~al.}(2017)\citenamefont{Stephens, Banigan,
  Adam, Goldman, and Marko}}]{Marko2017}
\bibinfo{author}{\bibfnamefont{A.~D.} \bibnamefont{Stephens}},
  \bibinfo{author}{\bibfnamefont{E.~J.} \bibnamefont{Banigan}},
  \bibinfo{author}{\bibfnamefont{S.~A.} \bibnamefont{Adam}},
  \bibinfo{author}{\bibfnamefont{R.~D.} \bibnamefont{Goldman}},
  \bibnamefont{and} \bibinfo{author}{\bibfnamefont{J.~F.} \bibnamefont{Marko}},
  \bibinfo{journal}{Molecular biology of the cell}
  \textbf{\bibinfo{volume}{28}}, \bibinfo{pages}{1984} (\bibinfo{year}{2017}).

\bibitem[{\citenamefont{Biggs et~al.}(2019)\citenamefont{Biggs, Liu, Stephens,
  and Marko}}]{Marko2019}
\bibinfo{author}{\bibfnamefont{R.}~\bibnamefont{Biggs}},
  \bibinfo{author}{\bibfnamefont{P.~Z.} \bibnamefont{Liu}},
  \bibinfo{author}{\bibfnamefont{A.~D.} \bibnamefont{Stephens}},
  \bibnamefont{and} \bibinfo{author}{\bibfnamefont{J.~F.} \bibnamefont{Marko}},
  \bibinfo{journal}{Molecular biology of the cell}
  \textbf{\bibinfo{volume}{30}}, \bibinfo{pages}{820} (\bibinfo{year}{2019}).

\bibitem[{\citenamefont{Irianto et~al.}(2013)\citenamefont{Irianto, Swift,
  Martins, McPhail, Knight, Discher, and Lee}}]{Discher2013}
\bibinfo{author}{\bibfnamefont{J.}~\bibnamefont{Irianto}},
  \bibinfo{author}{\bibfnamefont{J.}~\bibnamefont{Swift}},
  \bibinfo{author}{\bibfnamefont{R.~P.} \bibnamefont{Martins}},
  \bibinfo{author}{\bibfnamefont{G.~D.} \bibnamefont{McPhail}},
  \bibinfo{author}{\bibfnamefont{M.~M.} \bibnamefont{Knight}},
  \bibinfo{author}{\bibfnamefont{D.~E.} \bibnamefont{Discher}},
  \bibnamefont{and} \bibinfo{author}{\bibfnamefont{D.~A.} \bibnamefont{Lee}},
  \bibinfo{journal}{Biophysical journal} \textbf{\bibinfo{volume}{104}},
  \bibinfo{pages}{759} (\bibinfo{year}{2013}).

\bibitem[{\citenamefont{Naumova et~al.}(2013)\citenamefont{Naumova, Imakaev,
  Fudenberg, Zhan, Lajoie, Mirny, and Dekker}}]{Naumova2013}
\bibinfo{author}{\bibfnamefont{N.}~\bibnamefont{Naumova}},
  \bibinfo{author}{\bibfnamefont{M.}~\bibnamefont{Imakaev}},
  \bibinfo{author}{\bibfnamefont{G.}~\bibnamefont{Fudenberg}},
  \bibinfo{author}{\bibfnamefont{Y.}~\bibnamefont{Zhan}},
  \bibinfo{author}{\bibfnamefont{B.~R.} \bibnamefont{Lajoie}},
  \bibinfo{author}{\bibfnamefont{L.~A.} \bibnamefont{Mirny}}, \bibnamefont{and}
  \bibinfo{author}{\bibfnamefont{J.}~\bibnamefont{Dekker}},
  \bibinfo{journal}{Science (New York, N.Y.)} \textbf{\bibinfo{volume}{342}},
  \bibinfo{pages}{948} (\bibinfo{year}{2013}).

\bibitem[{\citenamefont{Ho et~al.}(2014)\citenamefont{Ho, Jung, Liu, Alver,
  Lee, Ikegami, Sohn, Minoda, Tolstorukov, Appert et~al.}}]{ho2014comparative}
\bibinfo{author}{\bibfnamefont{J.~W.} \bibnamefont{Ho}},
  \bibinfo{author}{\bibfnamefont{Y.~L.} \bibnamefont{Jung}},
  \bibinfo{author}{\bibfnamefont{T.}~\bibnamefont{Liu}},
  \bibinfo{author}{\bibfnamefont{B.~H.} \bibnamefont{Alver}},
  \bibinfo{author}{\bibfnamefont{S.}~\bibnamefont{Lee}},
  \bibinfo{author}{\bibfnamefont{K.}~\bibnamefont{Ikegami}},
  \bibinfo{author}{\bibfnamefont{K.-A.} \bibnamefont{Sohn}},
  \bibinfo{author}{\bibfnamefont{A.}~\bibnamefont{Minoda}},
  \bibinfo{author}{\bibfnamefont{M.~Y.} \bibnamefont{Tolstorukov}},
  \bibinfo{author}{\bibfnamefont{A.}~\bibnamefont{Appert}},
  \bibnamefont{et~al.}, \bibinfo{journal}{Nature}
  \textbf{\bibinfo{volume}{512}}, \bibinfo{pages}{449} (\bibinfo{year}{2014}).

\bibitem[{\citenamefont{Buxboim et~al.}(2017)\citenamefont{Buxboim, Irianto,
  Swift, Athirasala, Shin, Rehfeldt, and Discher}}]{buxboim2017coordinated}
\bibinfo{author}{\bibfnamefont{A.}~\bibnamefont{Buxboim}},
  \bibinfo{author}{\bibfnamefont{J.}~\bibnamefont{Irianto}},
  \bibinfo{author}{\bibfnamefont{J.}~\bibnamefont{Swift}},
  \bibinfo{author}{\bibfnamefont{A.}~\bibnamefont{Athirasala}},
  \bibinfo{author}{\bibfnamefont{J.-W.} \bibnamefont{Shin}},
  \bibinfo{author}{\bibfnamefont{F.}~\bibnamefont{Rehfeldt}}, \bibnamefont{and}
  \bibinfo{author}{\bibfnamefont{D.~E.} \bibnamefont{Discher}},
  \bibinfo{journal}{Molecular biology of the cell}
  \textbf{\bibinfo{volume}{28}}, \bibinfo{pages}{3333} (\bibinfo{year}{2017}).

\bibitem[{\citenamefont{Cho et~al.}(2017)\citenamefont{Cho, Irianto, and
  Discher}}]{cho2017mechanosensing}
\bibinfo{author}{\bibfnamefont{S.}~\bibnamefont{Cho}},
  \bibinfo{author}{\bibfnamefont{J.}~\bibnamefont{Irianto}}, \bibnamefont{and}
  \bibinfo{author}{\bibfnamefont{D.~E.} \bibnamefont{Discher}},
  \bibinfo{journal}{Journal of Cell Biology} \textbf{\bibinfo{volume}{216}},
  \bibinfo{pages}{305} (\bibinfo{year}{2017}).

\bibitem[{\citenamefont{Doyle and Underhill}(2005)}]{Doyle2005}
\bibinfo{author}{\bibfnamefont{P.~S.} \bibnamefont{Doyle}} \bibnamefont{and}
  \bibinfo{author}{\bibfnamefont{P.~T.} \bibnamefont{Underhill}},
  \emph{\bibinfo{title}{Brownian Dynamics Simulations of Polymers and Soft
  Matter}} (\bibinfo{publisher}{Springer Netherlands},
  \bibinfo{address}{Dordrecht}, \bibinfo{year}{2005}), pp.
  \bibinfo{pages}{2619--2630}.

\bibitem[{\citenamefont{Plimpton}(1995)}]{LAMMPS1995}
\bibinfo{author}{\bibfnamefont{S.}~\bibnamefont{Plimpton}},
  \bibinfo{journal}{Journal of Computational Physics}
  \textbf{\bibinfo{volume}{117}}, \bibinfo{pages}{1 } (\bibinfo{year}{1995}).

\bibitem[{\citenamefont{Guo et~al.}(2017)\citenamefont{Guo, Pegoraro, Mao,
  Zhou, Arany, Han, Burnette, Jensen, Kasza, Moore et~al.}}]{Guo2017}
\bibinfo{author}{\bibfnamefont{M.}~\bibnamefont{Guo}},
  \bibinfo{author}{\bibfnamefont{A.~F.} \bibnamefont{Pegoraro}},
  \bibinfo{author}{\bibfnamefont{A.}~\bibnamefont{Mao}},
  \bibinfo{author}{\bibfnamefont{E.~H.} \bibnamefont{Zhou}},
  \bibinfo{author}{\bibfnamefont{P.~R.} \bibnamefont{Arany}},
  \bibinfo{author}{\bibfnamefont{Y.}~\bibnamefont{Han}},
  \bibinfo{author}{\bibfnamefont{D.~T.} \bibnamefont{Burnette}},
  \bibinfo{author}{\bibfnamefont{M.~H.} \bibnamefont{Jensen}},
  \bibinfo{author}{\bibfnamefont{K.~E.} \bibnamefont{Kasza}},
  \bibinfo{author}{\bibfnamefont{J.~R.} \bibnamefont{Moore}},
  \bibnamefont{et~al.}, \bibinfo{journal}{Proceedings of the National Academy
  of Sciences of the United States of America} \textbf{\bibinfo{volume}{114}},
  \bibinfo{pages}{E8618} (\bibinfo{year}{2017}).

\bibitem[{\citenamefont{Adar and Safran}(2020)}]{Adar5604}
\bibinfo{author}{\bibfnamefont{R.~M.} \bibnamefont{Adar}} \bibnamefont{and}
  \bibinfo{author}{\bibfnamefont{S.~A.} \bibnamefont{Safran}},
  \bibinfo{journal}{Proceedings of the National Academy of Sciences}
  \textbf{\bibinfo{volume}{117}}, \bibinfo{pages}{5604} (\bibinfo{year}{2020}).

\end{thebibliography}
\end{document}